\documentclass[prl,aps,showpacs,twocolumn,superscriptaddress,amsmath,amssymb]{revtex4}
\usepackage{amsfonts}
\usepackage{amsmath}
\usepackage{graphicx}
\usepackage{dcolumn}
\usepackage{bm}

\newcommand*{\ket}[1]{\left| #1 \right>}

\begin{document}

\title{A versatile source of polarization-entangled photons}
\date{\today}

\author{A. Maser}
\email{Andreas.Maser@physik.uni-erlangen.de}
\affiliation{Institut f\"ur Optik, Information und Photonik, Universit\"at Erlangen-N\"urnberg, 91058 Erlangen, Germany}

\author{R. Wiegner}
\affiliation{Institut f\"ur Optik, Information und Photonik, Universit\"at Erlangen-N\"urnberg, 91058 Erlangen, Germany}

\author{U. Schilling}
\affiliation{Institut f\"ur Optik, Information und Photonik, Universit\"at Erlangen-N\"urnberg, 91058 Erlangen, Germany}



\author{C. Thiel}
\affiliation{Institut f\"ur Optik, Information und Photonik, Universit\"at Erlangen-N\"urnberg, 91058 Erlangen, Germany}


\author{J. von Zanthier}
\affiliation{Institut f\"ur Optik, Information und Photonik, Universit\"at Erlangen-N\"urnberg, 91058 Erlangen, Germany}



\begin{abstract}

We propose a method for the generation of a large variety of entangled states, encoded in the polarization degrees of freedom of $N$ photons, within the same experimental setup. Starting with uncorrelated photons, emitted from $N$ arbitrary single photon sources, and using linear optical tools only, we demonstrate the creation of all symmetric states, e.g., GHZ- and W-states, as well as all symmetric and non-symmetric total angular momentum eigenstates of the N qubit compound.

\end{abstract}

\maketitle

Entangled states, i.e., the nonseparable coherent superposition of multipartite states, play a key role in the investigations of fundamental aspects of quantum mechanics \cite{Mermin1990,Zeilinger1990}. They are also widely used as a basic resource for different tasks in quantum information processing \cite{Zeilinger2000a}, e.g., for applications in quantum cryptography \cite{Ekert1991,Gisin2002}, quantum teleportation \cite{Bennett1993}, or quantum computing \cite{Chuang2000,Briegel2001}. The experimental realization of a large variety of entangled states is thus highly desirable. 

The largest diversity of multipartite entangled states has been achieved so far by using the polarization degrees of freedom of photons, generated by spontaneous parametric down-conversion (SPDC) and subsequently fed into special arrangements of linear optical elements \cite{Zeilinger1999,Zeilinger2001,Weinfurter2003,Zeilinger2004,Steinberger2004,Zeilinger2004a,Guo2006,Pan2007,Weinfurter2007,White2008,Weinfurter2008,James2009,Langford2009,Weinfurter2009,Weinfurter2009a}. Although a large variety of different states can be produced by this technique, it usually suffers from the need of a particular experimental configuration for the generation of a particular entangled state \cite{Zeilinger1999,Zeilinger2001,Weinfurter2003,Zeilinger2004,Steinberger2004,Zeilinger2004a,Guo2006,Pan2007,Weinfurter2007,White2008,James2009}. Only very recently, experiments have been performed capable of creating more than one state out of the same entanglement class \cite{Langford2009}, a family of states \cite{Weinfurter2008}, or even different entanglement classes \cite{Weinfurter2007,Weinfurter2009,Weinfurter2009a}, inequivalent 
under stochastic local operations and classical communication (SLOCC) \cite{Cirac2000,Verstraete2002,Lamata2006}, within the same experimental setup. In the ideal case, however, one would  like to have a single apparatus that tunes in \textit{any wanted multipartite entangled state by simply turning a knob} \cite{Eisert2008}.

\begin{figure}
\begin{center}
\includegraphics[width=0.8\columnwidth]{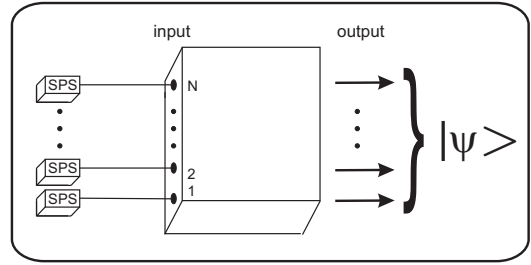}
\caption{The proposed method provides a black-box which can entangle $N$ uncorrelated single photons emitted from arbitrary single photon sources to a final state $\left|\psi\right>$.}
\label{fig:0}
\end{center}
\end{figure}

In this letter, we propose a technique, realizable with current technology, that allows to create an extremely large variety of multi-photon entangled states within the same experimental setup. This can be achieved by simply turning the orientation of polarization filters and/or extending the length of optical pathways, without changing otherwise the experimental design. The technique allows to generate for any number of photonic qubits all states symmetric under permutation of the photons, e.g., GHZ- and W-states, as well as the entire class of total angular momentum eigenstates of the multi-qubit compound. Amongst others, the scheme also enables to generate the canonical states representing all possible entanglement families of symmetric states inequivalent under SLOCC, as recently defined in \cite{Bastin2009a,Bastin2009b,Solano2009}. The method does not rely on a specific source but admits to employ arbitrary single photon sources emitting photons of identical frequency. The technique represents in this way a black-box capable of generating a large variety of entangled states starting from initially uncorrelated photons (see Fig.~\ref{fig:0}).


The proposed scheme consists of $N$ uncorrelated single photon sources emitting photons of identical frequency, e.g., trapped ions, neutral atoms, quantum dots, molecules or even photons produced via SPDC.
In front of each source a polarization filter is installed, which projects the polarization vector of the emitted photon onto the polarizer's axis $\bm{\epsilon}$. If the source emits photons of well defined polarization it suffices to turn the photon polarization vector along the desired axis $\bm{\epsilon}$ by use of a quarter- and a half-wave plate; the latter has the advantage that the count rate can be increased. Subsequently, the photons are registered by $N$ detectors located in the far-field region of the sources picking out $N$ spatial modes defined by the wave vectors $\vec{k}_n$, $n$ $\in 1, \ldots, N$. The far-field condition ensures that a given detector, upon detection, cannot distinguish which single photon source emitted the recorded photon, leading to a loss of Welcher-Weg information. 

\begin{figure}
\begin{center}
\includegraphics[width=0.8\columnwidth]{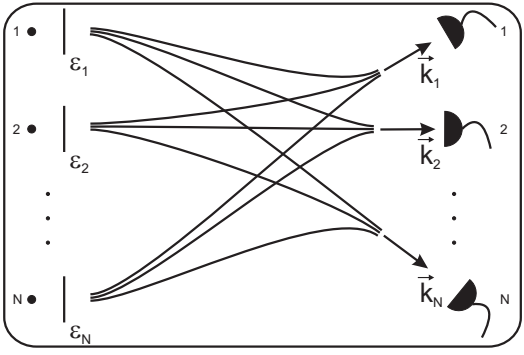}
\caption{The proposed setup implemented with optical monomode fibers. The black dots on the left hand side represent $N$ single photon sources each with a polarization device in front oriented along $\bm{\epsilon}$. The optical single mode fibers lead from each source to each detector.}
\label{fig:scheme}
\end{center}
\end{figure}

On their way from the sources to the detectors each photon will accumulate an optical phase $\phi_{n,m}$ given by

\begin{align}
\phi_{n,m}=k R_{n,m},
\label{phase}
\end{align}
where $R_{n,m}$ is the optical path length from source $n$ to detector $m$ ($n,m$ $\in 1, \ldots, N$) and $k$ is the wave number of the photons. In case of using optical fibers guiding the photons from the sources to the detectors \cite{Monroe2004,Weinfurter2006,Eschner2008} (see Fig.~\ref{fig:scheme}), $R_{n,m}$ corresponds to the optical path length through the fiber. By assuming that each detector registers exactly one photon, the correlated photon detection signal will display $N$-photon interferences.
By changing the orientation of the polarization filters/polarization turning device (called in the following \textit{polarization device}) and/or the optical phases $\phi_{n,m}$, a large variety of polarization-entangled photonic $N$-qubit states can be produced.

To see this in more detail let us start by considering a particular photon emitted by source $n$. Its wavefunction compatible with a successful measurement at any of the $N$ detectors is given by

\begin{align}
\left|\psi\right>&=\hat{\tilde{P}}_n\left|0,0,\ldots,0\right> \, ,
\end{align}
where
\begin{align}
\hat{\tilde{P}}_n&=\sum_{m=1}^N e^{i \phi_{n,m}} \left(\alpha_n \hat{a}_{\sigma^+}^{\dagger \left(m\right)}+ \beta_m \hat{a}_{\sigma^-}^{\dagger \left(m\right)}\right).
\end{align}
Hereby, the state vector $\left|\psi\right>=\left|x_1,x_2,\ldots,x_N\right> = \left|x_1\right>_1\otimes\left|x_2\right>_2 \otimes \ldots \otimes \left|x_N\right>_N$ describes the photon polarization state $\left|x_m\right>_m=\alpha_m \left| \sigma^+ \right>_m + \beta_m \left| \sigma^- \right>_m$ in the mode $\vec{k}_m$ for $m \in 1, \ldots, N$, $\hat{a}_{\sigma^\pm}^{\dagger \left(m\right)}$ is the creation operator of a photon with polarization $\sigma^\pm$ in mode $\vec{k}_m$, $\left|0,0,\ldots,0\right>$ is the vacuum state and the sum runs over all $N$ detectors as there is an equal probability for the photon to be registered by anyone of the $N$ detectors. The polarization vector of the photon after passing the polarization device in front of source $n$
is given by $\bm{\epsilon_n}=\alpha_n \bm{\sigma^+} + \beta_n \bm{\sigma^-}$, with arbitrary complex coefficients $\alpha_n$ and $\beta_n$ ($|\alpha_n|^2 + |\beta_n|^2 = 1$); the coefficient $\alpha_n$ ($\beta_n$) defines thus the probability amplitude of detecting a photon from source $n$ with polarization $\sigma^+$ ($\sigma^-$). 

To obtain the state of the system compatible with a successful detection event of $N$ photons we have to apply the $N$ operators $\hat{\tilde{P}}_n$, $n \in 1, \ldots, N$, onto the vacuum state $\left|0,0,\ldots,0\right>$, representing the emission of $N$ photons by the $N$ sources. However, since we consider only the case where each detector registers exactly one photon we have to change the operator $\hat{\tilde{P}}_n$ to $\hat{P}_n$ given by

\begin{align}
\hat{P}_n=\sum_{m=1}^N e^{i \phi_{n,m}} \left(\alpha_n \left|\sigma^+\right>_m\left<0\right|+ \beta_n \left|\sigma^-\right>_m\left<0\right|\right),
\label{emissionop}
\end{align}
where the operator $\left|\sigma^\pm\right>_m\left<0\right|$ creates a photon with polarization $\sigma^\pm$ in mode $\vec{k}_m$ iff the mode $\vec{k}_m$ was unpopulated before. In this way we ensure that each mode $\vec{k}_m$ is populated with no more than one photon.

The entire $N$ photon state compatible with a measurement of one photon in each mode 
is then given by

\begin{align}
\left|\psi_f\right>=\hat{P}_N\ldots\hat{P}_2 \hat{P}_1\left|0,0,...,0\right>.
\label{eq:photon}
\end{align}

To discuss the different possible outcomes for $\left|\psi_f\right>$ 
let us start by assuming that all optical  phases $\phi_{n,m}$ are multiples of $2\pi$. Then, the operators $\hat{P}_n$, $n \in 1, \ldots N$, and therefore also the final photonic state $\left|\psi_f\right>$, are totally symmetric under permutation of the modes. In this case $\left|\psi_f\right>$ can be expressed as a linear combination of the $N+1$ symmetric Dicke states with $k$ $\left|\sigma^-\right>$ excitations $\left|D_N\left(k\right)\right>$:

\begin{align}
\left|\psi_f\right>=\mathcal N \sum_{k=0}^N c_k \left|D_N\left(k\right)\right>,
\end{align}
where $\mathcal N$ is a normalization prefactor, and
\begin{align}
c_k=\binom{N}{k}^{\frac{1}{2}} \sum_{1 \leq i_1 \neq \ldots \neq i_N \leq N} \beta_{i_1} \ldots \beta_{i_k} \alpha_{i_{k+1}} \ldots \alpha_{i_N}.
\end{align}

To generate a particular symmetric Dicke state $\left|D_N\left(K\right)\right>$, $K \in 1, \ldots, N$, we have to choose the $\alpha_n$ and $\beta_n$ such that only $c_K\neq 0$, while all other $c_k=0$. This can be achieved by setting $\beta_1=\ldots=\beta_K=1$ and $\beta_{K+1}=\ldots=\beta_N=0$ (and therefore $\alpha_1=\ldots=\alpha_K=0$ and $\alpha_{K+1}=\ldots=\alpha_N=1$), i.e, by choosing the orientation of the polarization devices such that $K$ photons have polarization $\sigma^-$ and $(N-K)$ photons polarization $\sigma^+$ \cite{Thiel2007}. The symmetry of the setup then leads to a symmetric distribution of the photons among the modes.
  
This technique also allows to generate \textit{any} symmetric state $\left|\psi_{sym}\right>=\sum_{k=0}^N d_k\left|D_N\left(k\right)\right>$ with respect to permutations of the modes. For this, the polarization devices must be oriented such that for the polynomial of degree~$K$

\begin{align}
P\left(z\right)=\sum_{k=0}^N\left(-1\right)^{K-k}\sqrt{\binom{N}{k} / \binom{N}{K}} d_k z^k,
\label{polynom}
\end{align}
the $\alpha_n / \beta_n$ identify to the K roots of $P\left(z\right)$ for $n\leq K$, while the remaining $\beta_n$ ($\alpha_n$) are set to $0$ ($1$) \cite{Bastin2009}.

For the creation of non-symmetric states, we have to make full use of the degrees of freedom of the system, e.g., by varying the optical phases $\phi_{n,m}$ to values unequal to $2 \pi$ (e.g. by modifying the lengths of the optical fibers) or even removing optical fibers, i.e., not connecting all sources to all detectors. We used these degrees of freedom in former investigations to generate all (symmetric and non-symmetric) total angular momentum eigenstates among the two ground states $|\pm\rangle$ of $N$ trapped atoms with an internal $\Lambda$-level scheme \cite{Maser2009}. This can be achieved by detecting all photons emitted by the $N$ atoms on the transition from the excited state $|e\rangle$ to the ground states $|\pm\rangle$ by $N$ detectors located at particular positions in the far field. Examining the detection operator $\hat{D}_n$ which describes such a detection event \cite{Maser2009}

\begin{align}
\hat{D_n}=\sum_{m=1}^N e^{i \phi_{n,m}} \left(\alpha_n \left|-\right>_m\left<e\right|+ \beta_n \left|+\right>_m\left<e\right|\right),
\label{detectionop}
\end{align}
we find a striking similarity of $\hat{D}_n$ to our operator $\hat{P}_n$: in Eq.~(\ref{detectionop}), $\phi_{n,m}$ corresponds again to the optical phase accumulated by a photon when propagating from atom $n$ to detector $m$ (see Eq.~(\ref{phase})), and the operator $\left|\pm\right>_m\left<e\right|$ projects the previously excited atom $m$ to the ground state $\left|\pm\right>$. The final atomic state $\left|\psi_{atom}\right>$ is derived by applying the $N$ detection operators $\hat{D}_n$, $n \in 1, \ldots N$, to the initial atomic state $\left|e,e,\ldots,e\right>$, corresponding to the detection of the $N$ photons at the $N$ detectors

\begin{align}
\left|\psi_{atom}\right>=\hat{D}_N \ldots \hat{D}_1 \left|e,e,\ldots,e\right>.
\label{eq:ion}
\end{align}

By comparing Eq.~(\ref{detectionop}) with Eq.~(\ref{emissionop}), we can see that the $m$th atom accords to the $m$th mode in Fig.~\ref{fig:scheme}, an atom in the excited state $\left|e\right>$ to an empty mode $\left|0\right>$, and the two atomic ground states $\left|\pm\right>$ to the two photonic polarization states $\left|\sigma^\mp\right>$. Moreover, as Eq.~(\ref{eq:ion}) is formally equivalent to Eq.~(\ref{eq:photon}), this means that we can generate the same quantum states among the polarization degrees of freedom of $N$ photons as we can generate among the ground states of $N$ $\Lambda$-level atoms. In particular, if we allow to vary the optical phases $\phi_{n,m}$ to any value between $0$ and $2 \pi$ and/or to remove single optical fibers, suppressing thereby certain quantum paths, it becomes possible to generate any total angular momentum eigenstates of the $N$ photonic qubit compound, as demonstrated in case of atomic qubits in \cite{Maser2009}.

Using the notation of \cite{Maser2009} where the total angular momentum eigenstates are denoted by $\ket{S_1,S_2,S_3,\ldots,S_N;m_s}$ (where $S_1, S_2, ..., S_{N-1}$ takes the coupling history into account and $S_N$ and $m_N$ define the eigenvalues of the square of the total spin operator $\hat{\bf S}^2$ and its $z$-component $\hat{S}_z$, $S_N(S_N+1)\hbar^2$ and $m_N\hbar$, respectively~\cite{Dicke1954, Mandel1995}),
we can formulate the protocol to generate any of the photonic $N$-qubit total angular momentum eigenstates. In order to produce the state $\ket{S_1,S_2,S_3,\ldots,S_N;m_s}$ we have to 
\begin{itemize}
\item[1.] set up in front of the $N$ sources $\frac{N}{2}+m_s$ ($\frac{N}{2}-m_s$) $\sigma^-$ ($\sigma^+$) oriented polarization devices. Hereby, we connect all sources with optical fibers to the first detector.
\item[2.] check for each detector $j$ beginning with $j=2$ whether $S_j>S_{j-1}$ or $S_j<S_{j-1}$. If
\begin{itemize}
\item[a.] $S_j>S_{j-1}$, we have to connect detector $j$ with optical fibers to all sources except those which are mentioned in case [b.] below.
\item[b.] $S_j<S_{j-1}$, we have to connect detector $j$ with optical fibers to one source with a $\sigma^-$ polarization device and to one with a $\sigma^+$ polarization device in front. The optical fiber leading to the $\sigma^-$ polarization device should induce a relative optical phase shift of $\pi$ and those two sources should not be linked to any other subsequent detectors.
\end{itemize}
\end{itemize}
In this way, we can form any of the $2^N$ symmetric and non-symmetric total angular momentum eigenstates of the photonic $N$-qubit compound.

\begin{figure}
\begin{center}
\includegraphics[width=\columnwidth]{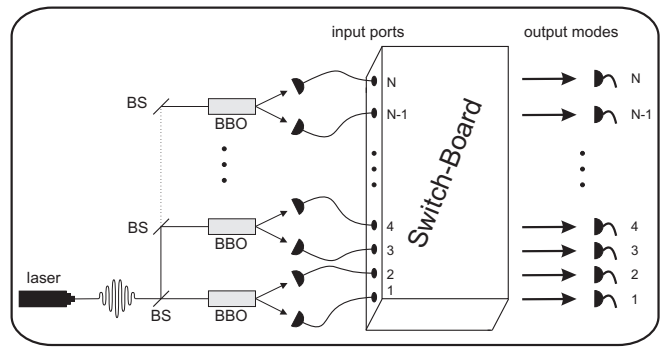}
\caption{Photonic multiqubit states generated by use of type-I SPDC as single photon sources and an appropriate configuration of optical fibers, considered as an optical switch-board.}
\label{fig:black_box}
\end{center}
\end{figure}

As SPDC is the brightest single photon source available it is favorable to employ it as a photon source for our proposed method. However, as mentioned, our method entangles $N$ \textit{uncorrelated} single photons. Therefore, in order to eliminate the momentum correlations among the photons generated via non-collinear SPDC, we can couple the photons into optical fibers. As each non-linear crystal scatters two SPDC photons, which can both be used, we need to set up $N/2$ non-linear crystals to supply $N$ single photons. Our method provides in this way an \textit{optical switch-board} which brings the initially uncorrelated photons into a large variety of entangled states (see Fig.~\ref{fig:black_box}). For the generation of a $4$ photon entangled state using two BBO crystals pumped by a pulsed Ti:Sapphire laser at $80$ MHz repetition rate, count rates of more than $4$ Hz can be expected \cite{White2008}.

In conclusion we proposed a scheme which allows to generate all symmetric states as well as all (symmetric and non-symmetric) total angular momentum eigenstates among $N$ photonic qubits within the same experimental setup. 
In particular, the scheme allows to generate all canonical states representing the possible entanglement families of symmetric states, inequivalent under SLOCC \cite{Bastin2009a,Bastin2009b}. To tune from one entanglement family of symmetric states to another, one just has to turn the orientation of the polarization devices in front of the sources. In this way, by appropriately configuring the optical fibers which connect the sources with the detectors, this optical switch-board allows to produce an extremely large variety of photonic multi-qubit states. We note that the protocol can be implemented by use of many different single photon sources, including photons generated by non-collinear SPDC. The latter has the advantage to obtain a significantly higher count rate compared to other single photon sources currently available. 

C. T. acknowledges financial support by the Staedtler Foundation.
R. W. gratefully acknowledge financial support by the Mayer Foundation and U.S. thanks the Elite Network of Bavaria for financial support.

\bibliographystyle{apsrev}
\bibliography{reference}

\end{document}